\documentclass[pra,twocolumn,showpacs,letterpaper,showpacs,superscriptaddress]{revtex4}
\usepackage{graphicx,amsmath,amssymb,amsfonts,latexsym,color,dcolumn,bm,subfigure,ulem}

\def\max{\mathrm{max}}
\def\min{\mathrm{min}}
\def\rms{\mathrm{rms}}
\def\res{\mathrm{res}}
\def\sp{\mathrm{sp}}
\def\ss{\mathrm{s,s}}
\def\pp{\mathrm{p,p}}
\def\s{\mathrm{s}}
\def\p{\mathrm{p}}

\begin{document}

\title{Kelvin probe force microscopy of metallic surfaces
used in Casimir force measurements}

\author{R. O. Behunin}
\affiliation{Department of Applied Physics, Yale University, New
Haven, Connecticut 06511, USA}

\author{D. A. R. Dalvit}
\affiliation{Theoretical Division, MS B213, Los Alamos National
Laboratory, Los Alamos, New Mexico 87545, USA}

\author{R. S. Decca}
\affiliation{Department of Physics, Indiana University-Purdue
University  Indianapolis, Indianapolis, Indiana 46202, USA}

\author{C. Genet}
\affiliation{ISIS \& icFRC, Universit\'{e} de Strasbourg and CNRS,
8, all\'{e}e Monge, 67000 Strasbourg, France}

\author{I. W. Jung}
\affiliation{Center for Nanoscale Materials, Argonne National
Laboratory, Argonne, Illinois 60439, USA}

\author{A. Lambrecht}
\affiliation{Laboratoire Kastler Brossel, UPMC-Sorbonne
Universit\'es, CNRS, ENS-PSL Research University, Coll\`ege de
France, Campus Jussieu, F-75252 Paris, France.}

\author{A. Liscio}
\affiliation{ISOF-CNR, via Gobetti 101, 40129 Bologna, Italy}

\author{D. L\'opez}
\affiliation{Center for Nanoscale Materials, Argonne National
Laboratory, Argonne, Illinois 60439, USA}

\author{S. Reynaud}
\affiliation{Laboratoire Kastler Brossel, UPMC-Sorbonne
Universit\'es, CNRS, ENS-PSL Research University, Coll\`ege de
France, Campus Jussieu, F-75252 Paris, France.}

\author{G. Schnoering}
\affiliation{ISIS \& icFRC, Universit\'{e} de Strasbourg and CNRS,
8, all\'{e}e Monge, 67000 Strasbourg, France}

\author{G. Voisin}
\affiliation{Department of Physics, Indiana University-Purdue
University  Indianapolis, Indianapolis, Indiana 46202, USA}
\affiliation{Laboratoire Univers et Th\'eories (LUTH), Observatoire de Paris; CNRS UMR8102,
Universit\'e Paris Diderot; 5 Place Jules Janssen, 92190 Meudon, France }

\author{Y. Zeng \footnote{Present address: Halliburton Energy Services, Houston, Texas 77032, USA}}
\affiliation{Theoretical Division, MS B213, Los Alamos National
Laboratory, Los Alamos, New Mexico 87545, USA}


\date{ \today}

\begin{abstract}

Kelvin probe force microscopy  at normal pressure was
performed by two different groups on the same Au-coated planar
sample used to measure the Casimir interaction in a sphere-plane
geometry. The obtained voltage distribution was used to calculate
the separation dependence of the electrostatic pressure $P_\res (D)$
in the configuration of the Casimir experiments. In the calculation
it was assumed that the potential distribution in the sphere has the
same statistical properties as the measured one, and that there are
no correlation effects on the potential distributions due to the
presence of the other surface. The result of this calculation, using
the currently available knowledge, is that $P_\res (D)$ does not
explain the magnitude or the separation dependence of the difference
$\Delta P (D)$ between the measured Casimir pressure and the one
calculated using a Drude model for the electromagnetic response of
Au. We discuss in the conclusions the points which have to be
checked out by future work, including the influence of pressure and
a more accurate determination of the patch distribution, in order to
confirm these results.

\end{abstract}


\maketitle

\section{Introduction}

Measurements of the Casimir interaction between gold-covered mirrors
now reach a good precision, which opens the way to detailed
comparisons with theoretical predictions. Some measurements,
performed at distances smaller than 1~$\mu$m, lead to unexpected
conclusions~\cite{Decca2005ap,Decca2007prd,Decca2007epj,Chang2012prb}.
These results agree with a description of conduction electrons in
metals by the lossless plasma model, and deviate significantly from
that based on the Drude model which accounts for
dissipation~\cite{Milton2005jpa,Klimchitskaya2009rmp,Ingold2009pre,Brevik2014ejp}.
Different conclusions are reached in another experiment performed at
distances of the order or larger than 1~$\mu$m~\cite{Sushkov2011nap}.
The results of this experiment agree with predictions drawn from the
dissipative Drude model, after the contribution of the electrostatic
patch effect has been subtracted.

In this context, it is important to discuss carefully all possible
sources of systematic effects, in particular the effect of
electrostatic patches already discussed for various high precision
measurements~\cite{Witteborn1967prl,Camp1991jap,Sandoghdar1996pra,%
Turchette2000prl,Deslauriers2006prl,Robertson2006cqg,Epstein2007pra,%
Pollack2008prl,Dubessy2009pra,Carter2011pra,Everitt2011prl,%
Reasenberg2011cqg,Hite2012prl,Hite2013mrs}, and more recently in the
context of Casimir force
measurements~\cite{Speake2003prl,Chumak2004prb,Kim2010pra,%
Kim2010jvst,Man2010jvst,Behunin2012pra,Behunin2012praa}. The patch
effect is due to the fact that the surface of a metallic plate is
made of micro-crystallites with different work
functions~\cite{Smoluchowski1941pr}. For clean metallic surfaces
studied by the techniques of surface physics, the resulting voltage
roughness is correlated to the grain size as well as to the
orientation of micro-crystallites~\cite{Gaillard2006apl}. For
surfaces exposed to air, the situation is changed due to the
unavoidable contamination by adsorbents, which spread out the
electrostatic patches, enlarge correlation lengths, and reduce
voltage dispersions~\cite{Rossi1992jap,Darling1992rmp,Leung2003prb}.

The force due to electrostatic patches can be computed by solving
the Poisson equation, as soon as the correlations of the patch
voltages are known. In other words, the force depends on the
associated voltage correlation function $C({\bf k})$,
with ${\bf k}$ a patch wavevector. In many studies devoted to this
question, the spectrum was assumed to be flat between two sharp
cutoffs at minimum and maximum wavevectors~\cite{Speake2003prl}.
Assuming that these cutoffs are given by the grain size distribution
measured with an atomic force microscope (AFM), it was concluded
that the patch pressure was much smaller than the difference between
the experimental Casimir pressure (more precise discussion below;
see Eq.\eqref{equivalentpressure}) and the theoretical prediction
based on the Drude model~\cite{Decca2005ap}.

A quasi-local model was proposed recently as a better motivated
representation of patches~\cite{Behunin2012pra}. The model produces
a smooth spectrum which leads to conclusions differing from those
drawn from the sharp-cutoff model, due to the contribution of low
values of $|{\bf k}|$. Using a very simple model with a uniform
distribution $\mathcal{P}(\ell)$ of patch sizes $\ell$ up to a
largest value $\ell^\max$ and a root-mean-square ({rms}) voltage
dispersion $V_\rms$, it was found that the difference $\Delta P (D)$
between experiment and theory based on the Drude model could be
qualitatively reproduced by fitting the model to the experimental
data. The corresponding values for $\ell ^\max$ and $V_\rms$ are
different from those obtained by identifying patch and grain sizes,
with $\ell ^\max\sim1~\mu$m larger than the maximum grain size
$\sim300$~nm, and $V_\rms\sim12$~mV smaller than the rms voltage
$\sim80$~mV associated with random orientations of clean
micro-crystallites of gold~\cite{Decca2005ap}. These values are
however compatible with a contamination of metallic surfaces, which
has to be expected
anyway~\cite{Rossi1992jap,Darling1992rmp,Leung2003prb}.

The results of~\cite{Behunin2012pra} imply that patches have to be
considered as an important source of systematic effects in Casimir
force measurements. However, they do not prove that patches are the
explanation of the difference $\Delta P (D)$ observed
in~\cite{Decca2005ap,Decca2007prd,Decca2007epj,Chang2012prb}. In
order to address this possibility, one has to measure the surface
voltage distribution on the samples used in Casimir experiments. The
method is to use the dedicated technique of Kelvin probe force
microscopy (KPFM) which has the ability of achieving the necessary
size and voltage
resolutions~\cite{NonnenmacherAPL1991,Liscio2008jpc,Liscio2008afm,Liscio2011acr}.
Using the measurements of patch potential distribution, it is then
possible to evaluate the contribution of the patches to the Casimir
measurements and to subtract it when comparing theory and
experiments. This evaluation has to be done in the plane-sphere
geometry by using results in~\cite{Behunin2012praa}.

The purpose of this paper is to present the first results of such an
analysis with measurements performed on the same Au-coated planar
sample  used to measure the Casimir interaction in a sphere-plane
geometry. The paper is organized as follows. In Section II we
briefly review Casimir measurements on gold samples performed at
Indiana University Purdue University Indianapolis (IUPUI). Section
III presents  normal pressure KPFM measurements of the same gold
samples. These measurements are carried out independently and
cross-checked in two separate laboratories, the one at IUPUI and
another one at Istituto per la Sintesi Organica e la
Fotoreattivit\'a (ISOF) in Bologna. We discuss the sample
preparation and characterization, as well as the measurement of the
patch properties. In Section IV we use the measured patch
distribution to compute the electrostatic interaction in the
sphere-plane geometry of Casimir experiments. As this is
experimentally more difficult, we have not performed KPFM
measurements on the spherical plates. We have instead used
properties demonstrated in~\cite{Behunin2012praa} to evaluate the
patch force by considering that the patch properties on the curved
surface are similar to those on the planar one. Within the
aforementioned caveats, the main conclusion of our study, discussed
in Section V, is that the calculated patch interaction does not have
the magnitude nor distance dependence which would explain the
difference $\Delta P (D)$  for the measurements reported
in~\cite{Decca2005ap,Decca2007prd,Decca2007epj}.


\section {Casimir effect measurements}

A planar  sample was made by sputtering 130 nm Au on a Si substrate.
Morphology and roughness studies performed by atomic force
microscopy  indicate excellent uniformity and low roughness on the
sample. The planar sample used in this paper is one of the many made
for the experiments reported in~\cite{grating}. The measured Casimir
interaction observed in this sample is indistinguishable within the
experimental error from the results reported
in~\cite{Decca2007prd,Decca2007epj}. The experimental setup for
measuring the Casimir effect is similar to the one used in previous
work~\cite{Decca2005ap,Decca2007prd,Decca2007epj}. A Au-coated
sapphire sphere (radius $R= (151.7 \pm 0.2)~\mu$m), is attached to a
micromechanical torsional oscillator. To enhance adhesion between
the $\sim$ 200 nm thick Au and the sapphire, a thin ($\sim$ 10 nm)
layer of Cr is first deposited on the sphere. The Au  layers in both
the sphere and the sample are thick enough to be considered infinite
from the Casimir interaction's stand point.

The sample is mounted on a flat platform which has an optical fiber
rigidly attached to it. The fiber axis coincides with the normal of
the sample-platform structure. The fiber is part of a two color
interferometer which keeps the sphere-sample separation $D$ stable
within half a nanometer. As the sample is brought into close
proximity of the sphere, the interaction between the two surfaces
produces a shift in the resonance frequency of the oscillator, which
is used to extract the gradient of the Casimir force, $\partial_D
F_C$. The use of a sphere instead of another planar surface avoids
the problem of keeping the two objects parallel but complicates the
exact theoretical description. A common approach to bypass this
difficulty relies on the proximity force approximation. When $D/R
\ll 1$,  one can then approximate the sphere's surface as a
collection of planar elements. Within this procedure, the force
gradient can be calculated as the sum of several local parallel
plane interactions, and
\begin{equation}
\partial_D F_C(D)=2\pi R P_{pp}(D)~, \label{equivalentpressure}
\end{equation}
where $P_{pp}(D)$ is the Lifshitz expression for the Casimir
pressure between two parallel plates~\cite{Lifshitz}.

As customarily done in Casimir force
measurements~\cite{Decca2005ap,Decca2007prd,Decca2007epj,Man2009pra,Decca2011ijmp},
the apparatus was calibrated using a calculable interaction, i.e.
the electrostatic interaction between the sphere and the sample. In
this section, we assume the two objects to be equipotentials, so
that the electrostatic energy between them is given by
\begin{equation}
E_e(D)=\frac{1}{2} C(D)\Delta V^2~, \label{energy}
\end{equation}
where $C(D)$ is the capacitance between the sphere and the plane
separated by a distance $D$, and the potential difference between
them $\Delta V = V_{\rm s}- V_{\rm p}$. An external voltage $V_{0}$
is applied between the two surfaces in the calibration
process~\cite{Decca2005ap,Decca2007prd,Decca2007epj,Man2009pra,Decca2011ijmp},
so that the potential difference becomes $\Delta V -V_{0}$.

From Eq.~\ref{energy} the force and the gradient of the force can be
easily derived when $\Delta V$ is not a function of distance. In the
calibration process both the expression of the force and the
gradient of the force have been used. It turns
out~\cite{Decca2005ap,Decca2007prd,Decca2007epj} that the force
\begin{equation}
F_e(D) =\frac{1}{2} \frac{\partial C(D)}{\partial D} \left(\Delta
V-V_{0}\right)^2 ~, \label{force}
\end{equation}
and the gradient of the force
\begin{equation}
\partial_D F_e(D) =\frac{1}{2} \frac{\partial^2 C(D)}{\partial D^2}
\left(\Delta V-V_{0}\right)^2 ~, \label{gradient}
\end{equation}
are not zero when $\Delta V =0$. With the simple Eqs.~\ref{force}
and~\ref{gradient} corresponding to equipotential surfaces, the
electrostatic interaction can be made null by a judicious choice of
the applied potential chosen to cancel the initial potential $V_0 =
V_{\rm min} = \Delta V$ ($V_{\rm min}$ is called the ``minimizing
potential"). A more precise discussion taking into account the patch
effect will be given below, in section IV.

\begin{figure}[tbp]
\centerline{\includegraphics[width=8cm]{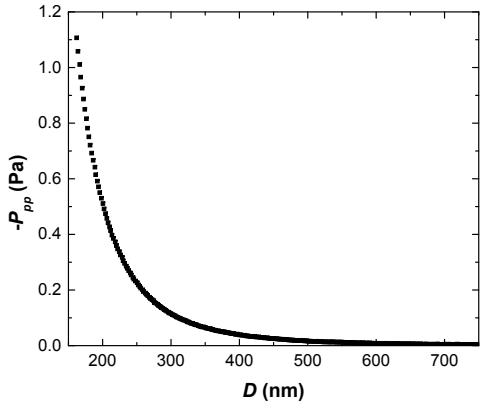}}
\caption{Equivalent Casimir pressure as a function of separation
between the sphere and the sample. Error bars in
$P$ ($\lesssim$ 3~mPa) and $D$ ($\sim$ 0.5~nm) are too small to be
seen.}
\label{figCasimir}
\end{figure}

In our calibration procedure, we have found that by taking the
derivative with respect to the potential difference, $V_{\rm min}$
is more accurately determined~\cite{Decca2009ijmpa}. Either the use
of the electrostatic force or the gradient of the electrostatic
force yield the same calibration parameters and, relevant for this
paper, the same value of  $V_{\rm min}$. This minimizing potential
was found to be independent of $D$ within the experimental accuracy
of 0.1 mV. The results of the Casimir interaction between the sphere
and the sample are shown in Fig.~\ref{figCasimir}.

\section{KPFM measurements}

The electrostatic potential distribution $V_{\rm p}({\bf x})$ on the
Au sample surface is measured by Kelvin probe force microscopy. This
contactless technique is based on monitoring long-ranged
electrostatic interactions between a cantilever and a sample. A
sharp metal-coated tip is microfabricated at the edge of a
cantilever which is maintained at a fixed potential. With no
mechanical action of the tip on the sample, electrostatic forces
exerted on the cantilever are measured, just as in AFM, by the
deflection of the cantilever using the reflection of a laser beam
off the tip~\cite{Liscio2011acr,MelitzSurfSci2011}. Because these
forces are proportional to the variation with distance $D$ of the
local capacitance $C$ between the tip and the sample, a direct
quantification of the surface potential difference $\Delta V$
between the tip and the sample is not trivial. To achieve this, KPFM
measurements exploit a Zeeman vibrating capacitor
setup~\cite{NonnenmacherAPL1991}. The two electrodes of the
capacitor are the sample and the tip which is forced to oscillate at
a fixed frequency $\omega$ while raster-scanning the surface of the
sample at fixed separation distance $D$. In such an amplitude
modulation (AM) mode, which is used for all KPFM measurements
reported in this paper, the tip oscillations modulate the tip-sample
electrostatic interaction energy $U(D)=\frac{1}{2} C \Delta V^2$,
assuming a linear relationship between local charges and local
potentials~\cite{JacobsJAP1998}. The electrical potential
inhomogeneities of the surface sample can thus be mapped by
detecting the amplitude variations of the free tip oscillations.

More precisely, a feedback loop applies an adjustable DC bias offset
potential $V_0$ to the cantilever tip in order to minimize the
interaction between the tip and the sample. Superimposed to this DC
voltage bias, an alternating current (AC) signal is applied to the
tip harmonically at a frequency $\omega$. In this case, $\Delta V$
is replaced in the expression for the interaction energy by the
total voltage $\Delta V -V_0+V_1 \sin (\omega t)$ between the tip
and the sample, where $V_1$ is the amplitude of the modulation.
Then, the $\omega$ component of the resulting force $F_\omega =
-\partial_D U_{\omega} = - \partial_D C \left[\left(\Delta V
-V_0\right)V_1 \sin (\omega t)\right]$, directly measured with a
lock-in amplifier, is canceled when $V_0=\Delta V$. The feedback
circuit monitors the bias $V_0$ applied to compensate for the
surface potential $\Delta V$, thus providing a direct quantification
of the latter. Note that the tip potential is calibrated using HOPG
(high ordered pyrolytic graphite), a substrate well stable in air.
This calibration implies that the real potential $V_{\rm p}({\bf
x})$ on the sample is determined up to a constant value (at a fixed
tip-sample distance). Such an offset does not affect the measurement
of the variations of the surface potential (see Section IV below for
a more precise discussion).

\begin{figure}[tbp]
\centerline{\includegraphics[width=6cm]{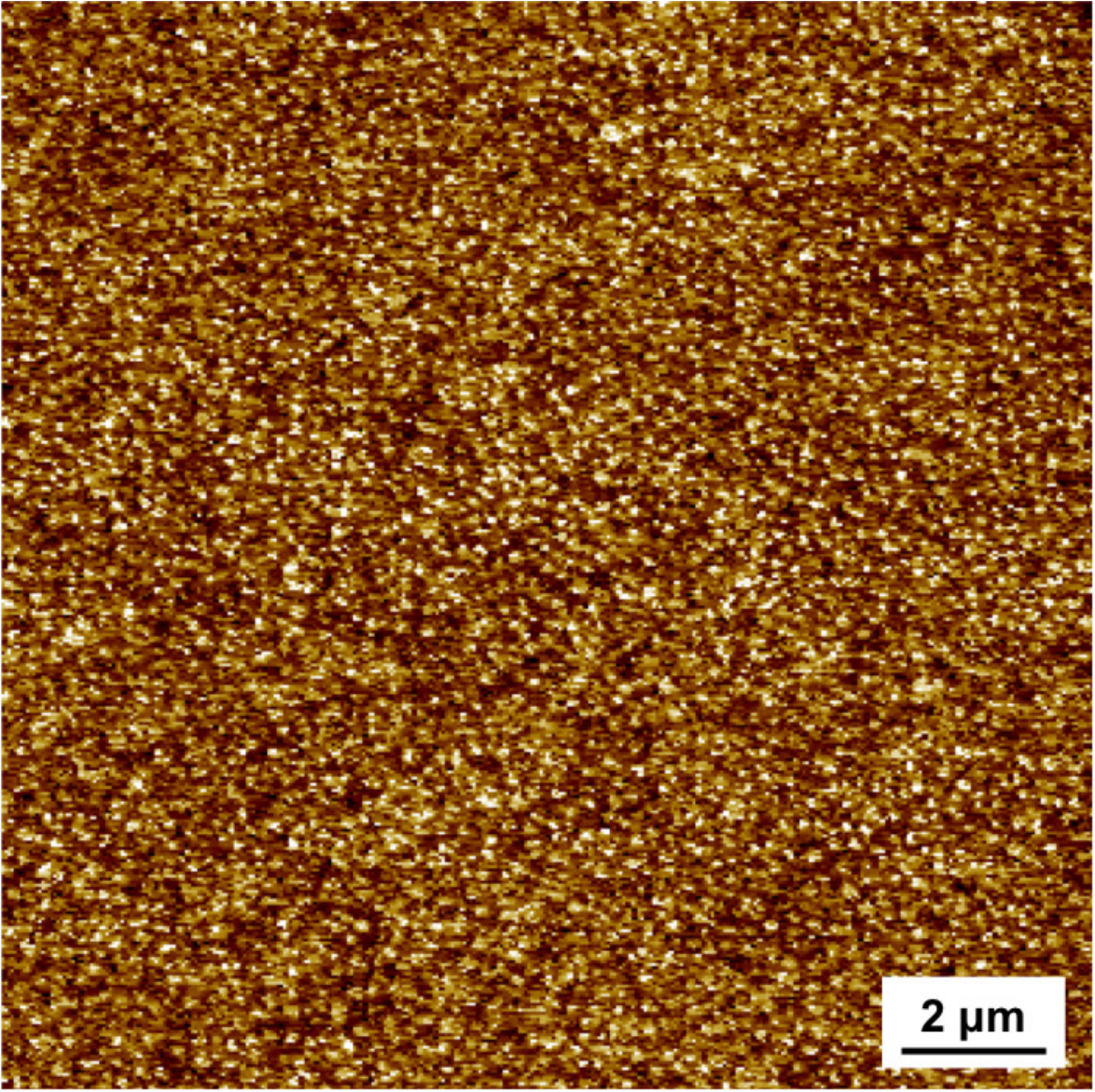}} \caption{
(Color online)  KPFM image of the electrostatic potential
distribution  $V_{\rm p}({\bf x})$ on the surface of the Au sample
recorded at ISOF. This image is composed of $512\times 512$ pixels,
with a lateral size of $15.36~\mu$m.The scale bar corresponds to
$2~\mu$m and the scan range is $20$ mV.  The amplitude of the
modulation is $V_1=2.5 V$. } \label{figBologna}
\end{figure}

The KPFM measurements shown in Fig. \ref{figBologna} have been
performed at ISOF using a commercial microscope Multimode III
(Bruker) equipped with an Extender Electronics module. The
measurements have been acquired in a nitrogen environment (relative
humidity smaller than $10\%$) at room temperature. Potential maps
have been recorded over a surface area of $15.36\times
15.36~\mu$m$^2$, with $512$ pixels per line, using a scanning rate
of $1$ Hz per line. In order to obtain a sufficiently large and
detectable mechanical deflection of the microscope tip, we used a
20~nm radius Pt/Ir coated Si ultra levers (SCM, Bruker) with
oscillating frequencies $\omega \sim (75 \pm 15)$ kHz and stiffness
$k\sim  2.8$ N$.$m$^{-1}$. The measurements have been performed at a
fixed tip-sample distance $ D=30$ nm, chosen as the minimal distance
that prevents artifacts due to the cross-talking between topographic
and electrical signals (precise criterium below).

Similar results were obtained at IUPUI using a different AFM
(Brueker Dimension)  with 20~nm radius Cr/Pt coated Si levers (Budget
Sensors, TAP190E-G) under similar environmental conditions. These
cantilevers are stiffer, with $k\sim  48$ N$.$m$^{-1}$ and a
resonance frequency $\sim (190 \pm 30)$ kHz. The KPFM measurements
were performed over a smaller area of 5 $\times$ 5~$\mu$m$^2$, with
256 points per line and at 1 Hz per line. The measurements were
repeated at different separations and it was found  that the
results from 20 nm to 60 nm were compatible and reproducible when
$V_1$ was kept below 3 V, without any cross-talks artifacts.

The criterium for the avoidance of cross-talking in both cases
(IUPUI and ISOF) is the observation of not too large correlations
$\langle h(x,y) \Delta V (x,y)\rangle <$ 0.5 between the height
$h(x,y)$ measured by the AFM at point $(x,y)$ and the potential
$\Delta V (x,y)$ measured by the KPFM at the same point.

Obviously, the measured  KPFM image is a convolution between the
real potential map and the microscope transfer function, leading to
unavoidable broadening of the nano-objects and underestimation of
the measured potential differences. The measured map can be
retrieved using linear deconvolution, although perfect recovery is
impossible without a precise description of the noise in the
system~\cite{MachleidtMST2009,CohenNano2013}. The transfer function
can be described in terms of tip-sample electrostatic interactions
and its width corresponds to effective surface area of the sample
interacting with the tip. Due to the long-range nature of the
electrostatic interactions, the area of the surface sampled in such
a measurement expands several tens of nanometers beyond the area
underneath the apex of the probe. In addition, the surrounding part
of the conical tip as well as the oscillating cantilever contribute
to the interaction. Other experimental parameters, in particular the
amplitude $V_1$ of the modulation, affect the transfer function and
its analytic evaluation requires a comprehensive simulation of the
tip-sample system~\cite{Charrier08}. Usually, the transfer function
has been calculated using simplified tip
geometries~\cite{Colchero01,Strassburg05}.

In this work, we have by-passed simulations and simplified
geometries by exploiting a semi-quantitative model developed in
Ref.~\cite{Liscio2011acr}. This approach has been checked by
measuring nano patterned samples with well-defined geometries.
Previous experiments performed with the same tip-sample geometry at
the same separation distance allowed us to evaluate an effective
microscope transfer function width of
$\sim$100 nm~\cite{Liscio2008jpc}. In the case of an isotropic
surface, the transfer function can be assumed to be Gaussian. In
this situation, a simple relation $w = 0.626\times L_R$ was recently
demonstrated between the width ($w$) of the effective area and
lateral resolution ($L_R$) defined as the minimal detectable feature
size~\cite{KalininNano2006,Liscio2008afm}. For ISOF measurements,
the pixel size of $30$~nm corresponds to a third of the effective
area width. This allows us to neglect pixelization and convolution
artifacts for areas larger than $160$~nm (i.e. larger than $5$
pixels width), and implies that the acquired KPFM images provide us
with fair maps of the gold surface potential for patch sizes larger
than $160$~nm. We tested this property by using the same KPFM
experimental setup to measure the electrical potential of
interdigitated gold nanoelectrodes having a channel length and an
electrode width of 200 nm. By comparing the applied potential and
the measured one, we observed an underestimation of the electrical
potential difference with AM-KPFM of the order of 20\%.


\section{Electrostatic patch interaction between a plane and a sphere}

In the following we recall the basic equations to evaluate the
electrostatic patch interaction between a plane and a sphere, using
the exact solutions derived in~\cite{Behunin2012praa}. In
particular, the known case of perfect equipotential surfaces on the
plane and the sphere can be solved in this way (see the Appendix C
in~\cite{Behunin2012praa}). Here we write the exact solutions for
patchy surfaces, and show how to deduce the patch interaction from
the KPFM data measured on the gold samples. Writing this interaction
as an equivalent pressure, as in Eq.\eqref{equivalentpressure}, we
finally compare our results to $\Delta P (D)$.

In order to solve the Poisson equation in the sphere-plane geometry
with arbitrary potential distributions on both surfaces, it is
advantageous to use bi-spherical coordinates because the equation is
then separable and the surfaces correspond to constant values of the
bi-spherical coordinate $\eta$~\cite{Behunin2012praa}. Writing the
boundary value problem for the electrostatic potential in the space
between the sphere and the plane, the interaction energy can be
expressed as a double integration over solid angles in bi-spherical
coordinates, $\int d\Omega\equiv \int_0^\pi d\xi \int_0^{2\pi} d\phi
\sin \xi$,  of a quadratic form of the surface potentials (see Eq.
(11) of~\cite{Behunin2012praa}). After performing a coordinate
transformation from bi-spherical to spherical or polar coordinates,
appropriate for the spherical and planar surfaces respectively, the
integration energy can be written in the form
\begin{equation}
E_\sp  = \sum_{a,b} \iint d\Omega_a d\Omega_b V_a(\Omega_a)
\mathcal{E}_{a,b}(\Omega_a;\Omega_b) V_b(\Omega_b) ~,
\label{exact-energy}
\end{equation}
where $V_{a,b}(\Omega_{a,b})$ denote the arbitrary electrostatic
potentials on the sphere and the plane (with $a,b=\s$ or $\p$
respectively), $\Omega_{\rm s}\equiv (\theta, \phi)$ are spherical
coordinates on the sphere, and $\Omega_{\rm p}\equiv (\rho,\phi)$
are polar coordinates on the plane. The integration measures are
defined as $\int d\Omega_{\rm s} = \int_0^{2 \pi} d\phi \int_0^{\pi}
d\theta \sin \theta$ (here $\theta$ is a polar angle on the sphere)
and $\int d\Omega_{\rm p}  = \int_0^{2 \pi} d\phi \int_0^{\infty}
d\rho \rho$ (here $\rho$ is the radius for a polar coordinate system
defined on the plane with origin below the apex of the sphere). The
kernels $\mathcal{E}_{a,b}(\Omega_a;\Omega_b)$ depend on the
distance $D$ between the sphere and the plane, and their explicit
expressions are given in Appendix B of~\cite{Behunin2012praa}. By
taking the derivative of the energy \eqref{exact-energy} with
respect to $D$, the electrostatic patch force between the sphere and
the plane is computed
\begin{eqnarray}
&&F_\sp  = \sum_{a,b} \iint d \Omega_a d\Omega_b V_a(\Omega_a)
\mathcal{F}_{a,b} (\Omega_a; \Omega_b) V_b(\Omega_b)
~, \nonumber\\
&&\mathcal{F}_{a,b} (\Omega_a; \Omega_b) =  \frac{\partial
\mathcal{E}_{a,b} (\Omega_a; \Omega_b)}{\partial D} ~.
\label{exact-force}
\end{eqnarray}
This expression is general for arbitrary boundary conditions on the
sphere and the plane.

As explained in Section III, we have measured the patch voltages on
the planar Au samples used in our Casimir force measurements, but we
do not have the same KPFM experimental knowledge for the sphere used
in Casimir experiments. In this context, we use the following
strategy to compute the total patch force. We consider that the
patch properties on the weakly curved surface ($R\gg D$) are similar
to those on the planar one on the length scales of relevance for our
calculation, and we use the fact known from~\cite{Behunin2012praa}
that the kernels $\mathcal{E}_\ss$ and $\mathcal{E}_\pp$ thus lead
to similar contributions (again for $R\gg D$). We also assume that
there are no statistical correlations between the patches on the
sphere and the plane ($\langle V_\s (\Omega_\s ) V_\p (\Omega_\p )
\rangle =0$), so that the kernel $\mathcal{E}_\mathrm{s,p}$ leads to
a negligible contribution. We then approximate the total force
between the plane and the sphere as twice the patch interaction
calculated in the simpler case when the sphere is grounded ($V_\s
=0$) and the plane has the patch distribution known from
measurements
\begin{equation}
\label{exact-force-approx} F_\sp  \approx  2  \iint d \Omega_{\rm p}
d\Omega_{\rm p}^\prime V_\p (\Omega_{\rm p}) \mathcal{F}_\pp
(\Omega_{\rm p}; \Omega_{\rm p}^\prime) V_\p (\Omega_{\rm
p}^\prime).
\end{equation}
We expect that this approximate expression for the patch force gives
the correct order of magnitude and distance dependence for the patch
interaction, provided the patch properties on the sphere and the
plane are similar, and the cross terms between the sphere and the
plane have a negligible contribution.

As discussed in Section II, an external voltage $V_{0}$ is applied
between the two surfaces in order to perform the electrostatic
calibration of the system. This bias $V_{0}$ is swept to observe the
quadratic dependence of \eqref{force} or \eqref{gradient} on $V_{0}$
at fixed sphere-plane separation $D$, and obtain its minimum which
defines the minimizing potential
\begin{equation}
\label{def-minimizing-potential} 0 = \left. \frac{\partial F_\sp
}{\partial V_0} \right|_{V_0=V_\min}  ~.
\end{equation}
A precise description of this problem is built up by adding a
constant value $V_0$ to the patchy potential $V_\p$ in
\eqref{exact-force-approx} and sweeping it. Solving
\eqref{def-minimizing-potential}, we find that $V_\min$ is defined
so that it compensates exactly the average value $\overline{V}_\p$
of the patch potential over the zone of electrostatic influence,
with the latter defined from the kernel
$\mathcal{F}_\pp$~\cite{Behunin2012praa}
\begin{eqnarray}
\label{minimizing-potential} &&V_\min = - \overline{V}_\p , \\
&&\overline{V}_\p \equiv \frac{ \int d\Omega_\p  \int d\Omega'_\p
V_\p (\Omega_\p ) \mathcal{F}_{\pp } (\Omega_\p ;\Omega'_\p)}
{\int d\Omega_\p \int d\Omega'_\p \mathcal{F}_{\pp }(\Omega_\p
;\Omega'_\p)} \nonumber ~.
\end{eqnarray}
The size of the zone of electrostatic influence is of the order of
$\sqrt{RD}\sim10~\mu$m, with the numbers corresponding to the
experiments in~\cite{Decca2005ap,Decca2007prd,Decca2007epj}. The
minimizing potential $V_\min$, which depends of the specific
realization of the patch voltage in the zone of electrostatic
influence, has to vary when the sphere-plane separation or the
lateral position of the sphere above the plane are changed. However,
this variation can be small due to the averaging of the effect of
patches over the zone of electrostatic influence.

With the more complete treatment of the electrostatic problem now
achieved, setting the applied potential $V_0$ equal to $V_\min$ does
no longer nullify the electrostatic interaction between the sphere
and the plane, but only minimizes it. There indeed remains the
effect of the dispersion of the patchy potential $V_\p$ over the
zone of electrostatic influence. This statement is made quantitative
by evaluating the residual patch force \eqref{exact-force-approx}
which remains at the minimizing potential
\eqref{minimizing-potential}
\begin{eqnarray}
\label{residual}
F_\res  &\equiv& \left. F_\sp \right|_{V_0=V_\min} \\
&=& 2 \iint d\Omega_{\rm p} d\Omega_{\rm p}^\prime \delta V_\p (\Omega_{\rm p})
\mathcal{F}_\pp (\Omega_\p;\Omega_\p^\prime) \delta V_\p (\Omega_{\rm p}^\prime)
~. \nonumber
\end{eqnarray}
Here $\delta V_\p(\Omega_{\rm p})$ is the deviation of the patchy
potential from its average over the zone of electrostatic influence
\begin{eqnarray}
\delta V_\p (\Omega_{\rm p}) \equiv V_\p (\Omega_\p) - \overline{V_\p} ~,
\end{eqnarray}
so that the residual patch force can effectively be
regarded as measuring the dispersion of $\delta V_\p(\Omega_\p)$ over the zone
of electrostatic influence.

At this point, it is worth discussing the contribution of patches
corresponding to given size scales. For small sizes, smaller than
the distance $D$ between the two plates, the contribution is
suppressed by the kernel $\mathcal{F}_\pp$ obtained by solving the
Poisson equation. For large sizes, larger than the size $\sqrt{RD}$
of the zone of electrostatic influence, the contribution could be
large before the calibration process, but it is essentially canceled
out in this process because $V_\min$ is defined so that it
compensates the average potential of patches over this zone. It
follows that the significant contributions are mainly associated to
size scales in the intermediate interval from $D$ to $\sqrt{RD}$,
that is from a fraction of $\mu$m to $10~\mu$m with the numbers
corresponding to the experiments
in~\cite{Decca2005ap,Decca2007prd,Decca2007epj}. These qualitative
statements are made precise by using the Eq.\eqref{residual}, with
the expression of the kernel $\mathcal{F}_\pp$ taken
from~\cite{Behunin2012praa}.

When performing numerical evaluations, we have to face the
difficulty that the measured samples are, of course, finite, as
discussed in Section III. In order to obtain patch distribution data
over a sufficiently large area, we used the following ``mirror
symmetry+replica" procedure. We took the measured KPFM data of the
finite-size square sample (we call it $1\times 1$ cell), generated a
$2\times 2$ cell by taking mirror images of the original $1\times 1$
cell, and then the $2 \times 2$ cell was periodically replicated on
two dimensions, until the final size reaches $80\times80~\mu{\rm
m}^2$, which is certainly enough for our numerics. Clearly, this
procedure introduces artificial correlations over distances larger
than the original sample sizes ($15\times15~\mu{\rm m}^2$ for the
larger ones), and it also ignores possible long-distance
correlations associated with very large patches. We believe our
method to be valid, at least for preliminary estimations, as a
consequence of the discussion of the preceding paragraph. The
contribution of possibly large patches (with sizes larger than
$\sqrt{RD}$) is essentially washed out in the electrostatic
calibration process because $V_\min$ compensates the average
potential of patches over the zone of electrostatic influence. We
computed the voltage correlation function from the KPFM data, and
the resulting correlation within the measurement area decreases as a
function of distance in an approximate exponential form. This
supports our assumption above for computing the electrostatic patch
interaction.

\begin{figure}[t]
\includegraphics[width=\columnwidth]{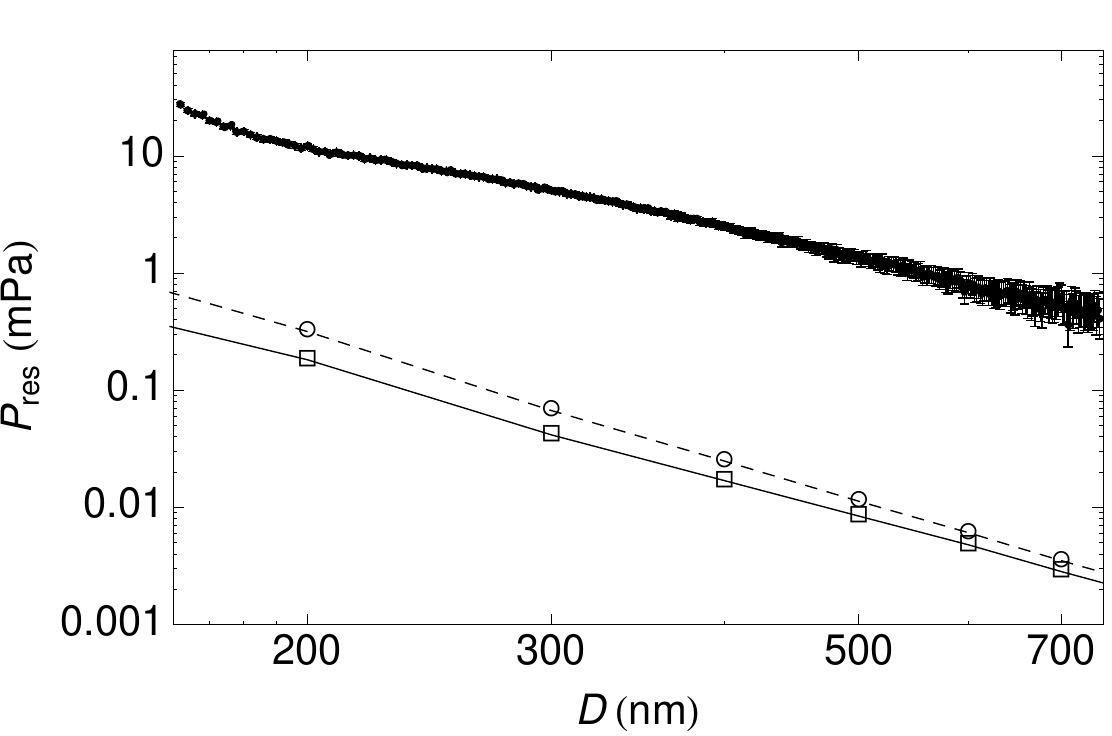}
\caption{Equivalent electrostatic patch pressure $P_\res$ computed
for the IUPUI (solid line with squares) and ISOF (dashed line with
circles)  data, versus distance $D$. We also show, for comparison,
the difference $\Delta P (D)$  between experimental measurements of
the Casimir pressure and theoretical predictions based on the Drude
model.} \label{figpatchpressure}
\end{figure}

Fig.\ref{figpatchpressure} shows our numerical results for the patch
interaction, measured as an equivalent patch pressure as in
Eq.\eqref{equivalentpressure}. Though they were obtained on
different parts of the same sample with different instruments, scan
sizes and resolutions, the measurements made at IUPUI (solid line
with squares) and ISOF (dashed line with circles) lead to comparable
patch pressures, in terms of their magnitude and variation with
distance. In particular, both curves have a different law of
variation with $D$ and smaller magnitudes  than $\Delta P (D)$, also
reproduced for comparison on Fig.\ref{figpatchpressure}. For this
difference, the bars show the experimental uncertainties discussed
in~\cite{Decca2007prd,Decca2007epj}, similar to those shown in
Fig.\ref{figCasimir}. The theoretical calculations from the Drude
model are described
in~\cite{Decca2005ap,Decca2007prd,Behunin2012pra}. They are done at
room temperature $T=295$~K using tabulated optical data extrapolated
to low frequencies with a Drude model with parameters $\Omega_{\rm
P}=8.9$~eV for the plasma frequency and $\gamma = 0.0357$~eV for the
damping rate. A simple model for roughness corrections is
used~\cite{Decca2007prd}, with root-mean square roughness heights
of 3.6~nm and 1.9~nm, for the plane and the sphere respectively.


\section{Conclusions}

In this paper, we have shown that it is possible to measure patch
properties on the same Au samples used in Casimir
experiments~\cite{Decca2007prd,Decca2007epj}. In fact we did it on
the planar samples and we assumed that the properties were similar
on the spherical ones. We then estimated the contribution of patches
to the force in the plane-sphere geometry used in Casimir
experiments~\cite{Behunin2012praa}.

We have discussed the subtleties associated to small and large patch
sizes. The influence of patch sizes smaller than the plane-sphere
distance $D$ is suppressed in the solution of the Poisson equation.
The influence of patch areas larger than the zone of influence $2\pi
R D$ is canceled by the voltage $V_0$ applied in the electrostatic
calibration. This entails that, for the parameters used in the
Casimir experiment, the significant contributions from patches are
mainly associated to sizes in the interval from a fraction of a
$\mu$m to $\sim$10~$\mu$m. Hence the resolution of the AM-KPFM
measurements discussed in Section III should be sufficient for a
reliable estimation of the effect of electrostatic patches shown in
Section IV.

The patch pressure estimations shown in Fig.\ref{figpatchpressure}
have smaller magnitudes and a different law of variation with $D$
than the difference $\Delta P (D)$ observed in Casimir
experiments~\cite{Decca2007prd,Decca2007epj}. They do not reproduce
the results which were found in~\cite{Behunin2012pra} to fit this
difference. This means that the statistical properties measured on
the patches differ from the model used in~\cite{Behunin2012pra}. It
has also to be emphasized at this point that the description of the
patch interaction in~\cite{Behunin2012pra} was based on the
proximity force approximation, whereas the present paper used the
much more satisfactory approach developed in~\cite{Behunin2012praa}
to perform precise evaluations in the plane-sphere geometry.

The analysis of the present paper is preliminary and some of its
limitations have to be cured by further work. In our calculation of
the sphere-plane patch force \eqref{exact-force-approx}, we have
assumed that the patches on the sphere had the same statistical
properties as on the plane, and also that the cross-correlations
between the patches on the sphere and plane had a negligible
contribution. In order to confirm these assumptions, it would be
necessary to measure patches on the spherical mirrors, which is an
experimental challenge. We have measured patch distributions on
samples at ambient pressures, whereas the Casimir experiments were
performed at $\sim 10^{-7}$~torr. As the pressure could influence
the contamination process and hence the patch properties, it would
be crucial to repeat the patch characterization on the same metallic
samples and under the same environmental conditions as in the vessel
where Casimir measurements are done.

Our KPFM measurements were done with a scan size of the order of
$15~\mu$m and a resolution of the order of $160$nm (ISOF
experiment). Such numbers should be sufficient to get a qualitative
characterization, as they cover the patch sizes having a critical
influence on the force between the plane and the sphere. Of course,
larger scan sizes and improved resolutions would allow one to test
the reliability and accuracy of the whole method. Further work is
thus needed to confirm the present result that the patch
contribution does not match the difference $\Delta P (D)$ observed
in Casimir
experiments~\cite{Decca2005ap,Decca2007prd,Decca2007epj,Chang2012prb}.

{\it Note added:} While this paper was under review, a
preprint~\cite{Munday2014} has become available with conclusions
differing from ours. It is asserted there that AM-KPFM measurements
underestimate the potential differences as measured by FM-KPFM and
thus lead only to a lower bound for the patch contribution to the
force. Underestimations by AM-KPFM of the true potential differences
on metallic samples have indeed been reported~\cite{Zerweck2005},
and they depend on experimental conditions. As explained at the end
of \S~III, we have checked that the underestimation is of the order
of 20\% for typical patch sizes of 200 nm and under the experimental
conditions used in our measurements. Though it calls for further
work in order to confirm the results of the present paper, such an
underestimation does not affect its conclusions (see the preceding
paragraphs).


\subsection*{Acknowledgments}

The authors thank the participants to the ESF Research Networking
Programme CASIMIR (www.casimirnetwork.com) for many discussions
related to the topic of this paper. Work at Los Alamos National
Laboratory was carried out under the auspices of the NNSA of the
U.S. DOE under Award No. DEAC52-06NA25396 and the LANL LDRD program.  
This work was performed, in part, at the Center for Nanoscale Materials, a U.S. Department of
Energy, Office of Science, Office of Basic Energy Sciences User
Facility under Contract No. DE-AC02-06CH11357. R.S.D. acknowledges
support from the IUPUI Nanoscale Imaging Center, Integrated
Nanosystems Development Institute, Indiana University Collaborative
Research Grants and the Indiana University Center for Space
Symmetries.

\end{document}